# High-Frequency Modeling and Simulation of a Single-Phase Three-Winding Transformer Including Taps in Regulating Winding

(to be submitted to IEEE Trans. Power Delivery)

Bjørn Gustavsen, *Fellow*, *IEEE*, Alvaro Portillo, *Senior Member*, *IEEE,*
Rodrigo Ronchi, *Member*, *IEEE,* and Asgeir Mjelve, *Member*, *IEEE*

*Abstract*—Transformer terminal equivalents obtained via admittance measurements are suitable for simulating high-frequency transient interaction between the transformer and the network. This paper augments the terminal equivalent approach with a measurement-based voltage transfer function model which permits calculation of voltages at internal points in the regulating winding. The approach is demonstrated for a single-phase three-winding transformer in tap position Nom+ with inclusion of three internal points in the regulating winding that represent the mid-point and the two extreme ends. The terminal equivalent modeling makes use of additional common-mode measurements to avoid error magnifications to result from the ungrounded tertiary winding. The final model is used in a time domain simulation where ground-fault initiation results in a resonant voltage build-up in the winding. It is shown that that the peak value of the resonant overvoltage can be higher than during the lightning impulse test, with unfavorable network conditions. Additional measurements show that the selected tap position affects the terminal behavior of the transformer, changing the frequency and peak value of the lower resonance point in the voltage transfer between windings.

## I. Introduction

TRANSIENT overvoltages are one of the root-causes of dielectric failures in transformers. The assessment of internal winding dielectric stresses is in the case of power transformers routinely performed by the manufacturers using in-house calculation programs. These programs make use of a detailed description of the transformer windings with circuit element parameters obtained using analytical expressions or Finite Element calculations, see [1] and references therein for an overview of methods. It has been shown [2] that when applied to a common geometry, the calculation programs by different manufacturers generally give similar results for the peak value of external and internal voltages when applying a lightning impulse voltage to the transformer. However, substantial differences were found for the resonance frequencies and damping. This fact raises concerns about the suitability of the manufacturer's models when applied in simulations involving general overvoltages that result from the connected system. On the other hand, measurement-based black-box models can provide quite accurate results [3]-[7] for the voltage on external terminals, although the transformer must first be built before any measurements can be performed.

One of the dielectric weak parts of a transformer is the regulating winding which, depending on the selected tap position, can exhibit strong internal resonances. In the 1970s, American Electric Power experienced several failures in auto-transformer regulating windings [8], with resonant voltage build-up as root-cause. It is therefore of interest to extend the capability of the black-box model approach to also include points in the regulating winding.

CIGRE JWG A2/C4.52 has in 2016 performed a measurement campaign on a single-phase and a three-phase transformer in order to assess/improve the accuracy of currently applied white-box models, and to provide input for black-box and grey-box modeling. The measurements involve frequency domain and time domain measurements at the transformer external terminals and at some internal points.

This paper reports results obtained for black-box-modeling of the single-phase three-winding transformer. In addition to the four external terminals, the measurements include voltage transfer from external terminals to three points in the regulating winding. From the measurements, a wide-band model is extracted which can be applied in transient simulation in an EMTP-type environment. The various steps in the measurement and modeling procedure are described with emphasis on accuracy preservation. A few comparisons are made with white-box model simulation results to appreciate the fidelity of the proposed model. Also, the effect of the tap changer setting on transferred overvoltages between the windings is investigated. Finally, the model is applied in a transient simulation where a resonant voltage build-up takes place in the regulating winding. The resulting overvoltages are compared with those arising in the standard lightning impulse test.

## II. Transformer Unit And Measured Quantities

The transformer is a 50 MVA single-phase three-winding transformer with rated voltages 230/69/13.8 kV at 60 Hz. The core has one mid-leg and two return legs. The internal connections are shown in Fig. 1. The regulating winding is a an interleaved disk winding with 10 circuits. The online tap changer (OLTC) has 11 tap positions and the polarity of this winding is reversible, giving a total of 21 tap positions.

The measurements were performed with the transformer active part inside the tank, but without oil, lid and bushings. This paper shows results for measurements with the tap changer in mid position at position Nom+, i.e. with the moving contact "k" in Fig. 1 in position "+" and with the moving contact H0 in R11. It is observed that in this tap setting, R1 has no galvanic connection to other parts in the transformer, thereby being susceptible to voltage oscillations and therefore vulnerable to overvoltages.

The external connection points H1, H0, X1, X0, Y1 and Y2 were brought to the rim of the tank (Fig. 2) where they were made available for the measurement equipment which was placed on top of a scaffolding. The applied connection leads were those already installed for connecting H1 and X1 to the bushings as well as additional leads for bringing the remaining connection points to the rim. Three additional measurement leads were installed that allowed measurement at three internal points in the regulating winding: the two extreme ends (R1, R11) and a point (R5) near the mid-point (R6). A braided wire was clamped to the tank rim and used as ground reference in the measurements.

Using a setup similar to the one in [4], the admittance matrix was measured with respect to the four external terminals H1, X1, Y1, Y2, with terminals H0 and X0 grounded at the tank rim. In addition, two identical voltage probes were used for measuring voltage transfer functions from these four external terminals to the three points in the regulating winding, R1, R5, and R11.

Tables I and II define the node labeling used in this work as well as the type of measurements that were performed.

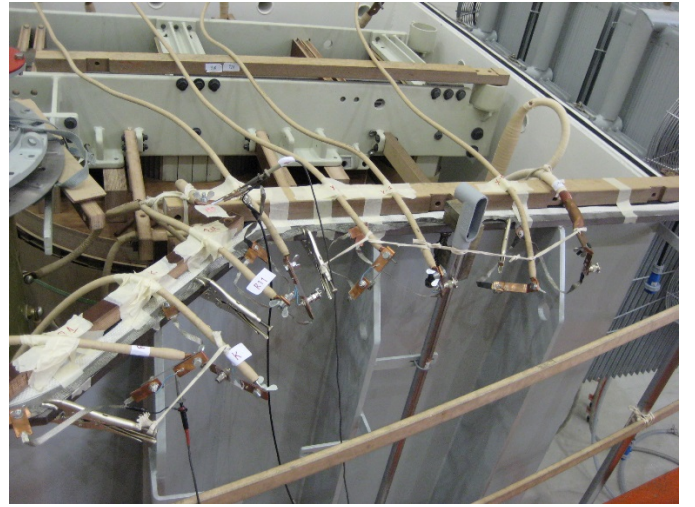

Fig. 2. Measurement connections.

TABLE I EXTERNAL NODES (TERMINALS)

| Node | Terminal | Description | Measurement type |
|---|---|---|---|
| 1 | H1 | HV winding | Admittance |
| 2 | X1 | LV winding | Admittance |
| 3 | Y1 | Tertiary winding. | Admittance |
| 4 | Y2 | Tertiary winding. | Admittance |

TABLE II INTERNAL NODES

| Node | Terminal | Description | Measurement type |
|---|---|---|---|
| 5 | R1 | Tap position 1 | Voltage transfer |
| 6 | R5 | Tap position 5 | Voltage transfer |
| 7 | R11 | Tap position 11 | Voltage transfer |

## III. BLACK-BOX MODELING USING COMBINED ADMITTANCE AND VOLTAGE TRANSFER MODELING

In principle, one can create a terminal admittance model with respect to all seven terminals. However, since nothing will be connected to the internal points, we chose a different strategy [9] which uses a combination of a terminal model with respect to the external terminals only (H1, X1, Y1, Y2) with the voltage on internal points obtained via voltage transfer functions. This latter approach has the advantage of fewer measurements and less difficulties with passivity enforcement since there are no passivity requirements for the voltage transfer model. The procedure is outlined below and described in detail in Sections IV and V.

The input for the modeling is a measurement of the terminal admittance matrix $\mathbf{Y}(\omega)$ and the voltage transfer matrix $\mathbf{H}(\omega)$ from external terminals to internal points, see Fig. 3. $\mathbf{Y}$ relates the voltages and currents at the transformer external terminals (1) while $\mathbf{H}$ determines the internal voltages with the external voltages as input (2).

$$\mathbf{i}_{ext}^{4\times 1}(\omega) = \mathbf{Y}^{4\times 4}(\omega)\, \mathbf{v}_{ext}^{4\times 1}(\omega) \qquad (1)$$

$$\mathbf{v}_{int}^{3\times 1}(\omega) = \mathbf{H}^{3\times 4}(\omega)\, \mathbf{v}_{ext}^{4\times 1}(\omega) \qquad (2)$$

Matrices $\mathbf{Y}$ and $\mathbf{H}$ are to be fitted with multi-port rational functions on state-space form,

$$\mathbf{Y}(\omega) = \mathbf{C}^Y (s\mathbf{I} - \mathbf{A}^Y)^{-1} \mathbf{B}^Y + \mathbf{D}^Y \qquad (3)$$

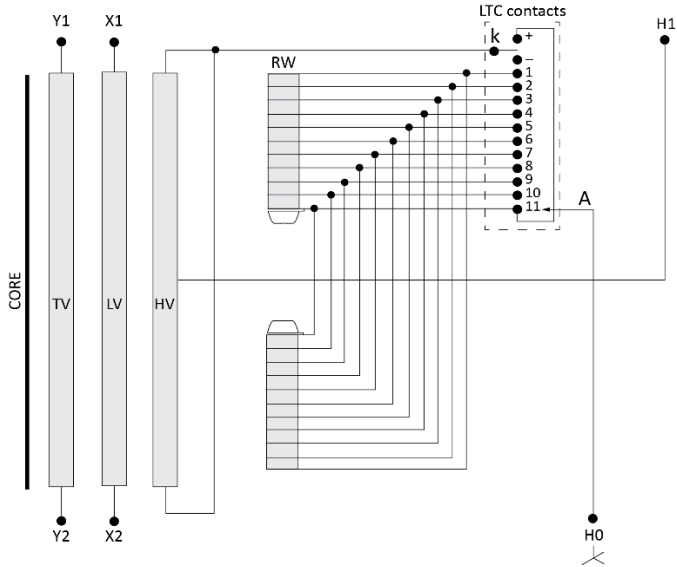

Fig. 1. Single-phase transformer.

$$\mathbf{H}(\omega) \cong \mathbf{C}^H (s\mathbf{I} - \mathbf{A}^H)^{-1} \mathbf{B}^H + \mathbf{D}^H \quad (4)$$

The two state-space models (3) and (4) are utilized in an EMTP-type time domain simulation environment as shown in Fig. 4. The admittance model is connected to the power system and the simulation gives the voltages at the transformer external terminals. These (four) voltages are used as input to the voltage transfer block whose outputs are the voltage waveforms at the (three) internal nodes.

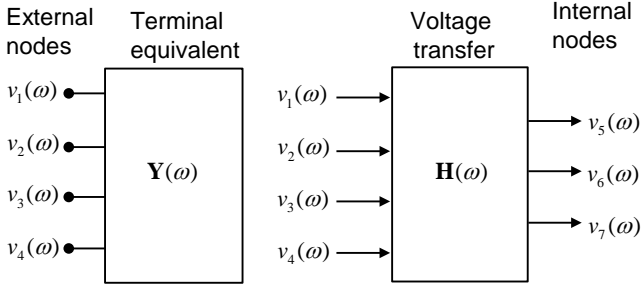

Fig. 3. External and internal nodes

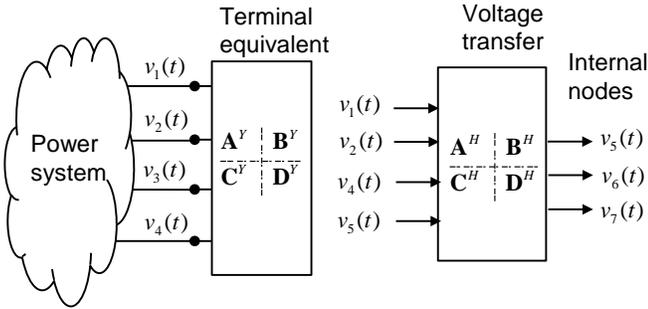

Fig. 4. Inclusion of model in time domain simulation.

## IV. ADMITTANCE MODELING

### A. Measurements

Fig. 5 shows the measured elements of the admittance matrix $\mathbf{Y}$ with respect to the four external terminals. It is observed that the four elements that are related to the tertiary winding (terminals 3 and 4) are practically equal in magnitude at frequencies below 10 kHz. This result is due to the tertiary winding being ungrounded which causes the common mode component, which corresponds to the capacitive charging current, to approach zero as frequency is reduced. This behavior is however inaccurately represented in the short-circuit measurement which leads to error magnifications in calculations of voltage transfers to the tertiary winding. This problem is demonstrated in Fig. 6 where the voltage transfer from terminal 2 (X1) to open terminals 1 (H1), 3 (Y1) and 4 (Y2) has been calculated from the measured $\mathbf{Y}$. It is seen that excessive errors arise a frequencies below 10 kHz. Additional errors exist at high er frequencies as well but they have a different cause which is eliminated in Section IV-B.

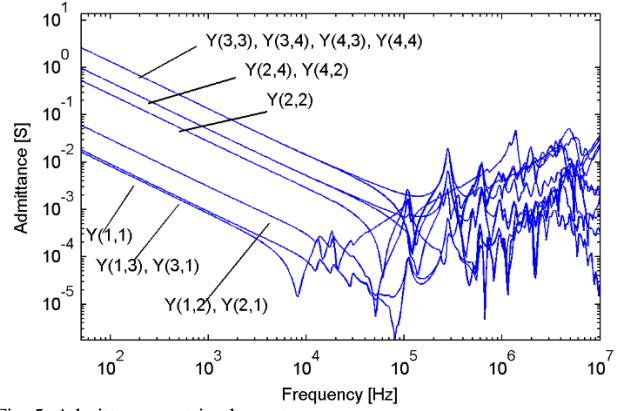

Fig. 5. Admittance matrix elements.

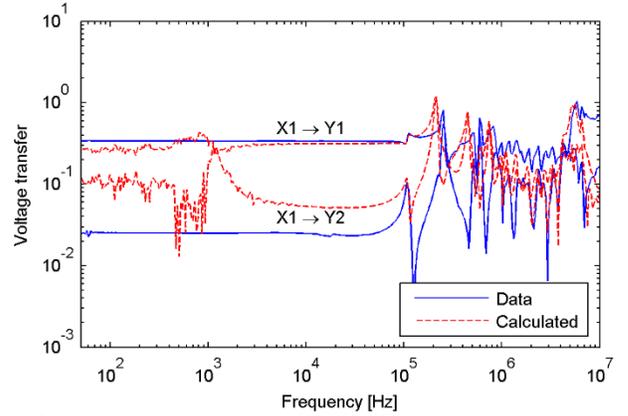

Fig. 6. Voltage transfer from X1 to Y1 and Y2, with H1 open.

In order to overcome the accuracy limitations, we will explicitly model the common mode components. Consider the matrix partitioning of $\mathbf{Y}$ in (5). As already mentioned, the 2×2 block associated with Y1 and Y2 has a very small (common mode) eigenvalue. It can be further shown that the coupling from H1 and X1 to Y1 and Y2 will also have a very small common mode component at low frequencies, which must be accurately represented.

$$\mathbf{Y} = \begin{bmatrix} Y(1,1) & Y(1,2) & Y(1,3) & Y(1,4) \\ Y(2,1) & Y(2,2) & Y(2,3) & Y(2,4) \\ Y(3,1) & Y(3,2) & Y(3,3) & Y(3,4) \\ Y(4,1) & Y(4,2) & Y(4,3) & Y(4,4) \end{bmatrix} \begin{matrix} \text{H1} \\ \text{X1} \\ \text{Y1} \\ \text{Y2} \end{matrix} \quad (5)$$

(columns: H1, X1, Y1, Y2)

Additional measurements were performed with terminals 3 and 4 bonded, see Fig. 7. The admittance measurement of the resulting 3×3 matrix $\mathbf{Y}'$ now directly reveals the common mode component associated with Y1 and Y2.

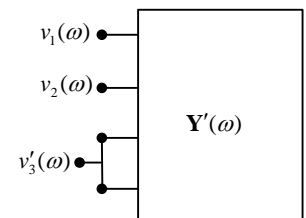

Fig. 7. Common mode admittance measurements.

Fig. 8 shows elements $Y'(3,3)$, $Y'(3,1)$ and $Y'(3,2)$. The elements at frequencies below 10 kHz are replaced with the expected capacitive behavior (straight lines) that was deduced from the 10 kHz sample point. These measurements were next used for replacing the common mode behavior of the original 4×4 **Y** as follows.

1. For the 2×2 block associated with terminals 3 and 4, subtract from each row the average of the row-sum, and from each column the average of the column sum. Add to all four elements the measured element $Y'(3,3)$ divided by four.
2. Subtract from elements $Y(3,1)$ and $Y(4,1)$ the average of the two elements, and add to the two elements the measured $Y'(3,1)$ divided by two. Copy the two modified elements into position $Y(1,3)$ and $Y(1,4)$.
3. Subtract from elements $Y(3,2)$ and $Y(4,2)$ the average of the two elements, and add to the two elements the measured $Y'(3,2)$ divided by two. Copy the two modified elements into position $Y(2,3)$ and $Y(2,4)$.

The above replacement is only made at frequencies below 10 kHz as the accuracy was otherwise found to deteriorate at higher frequencies.

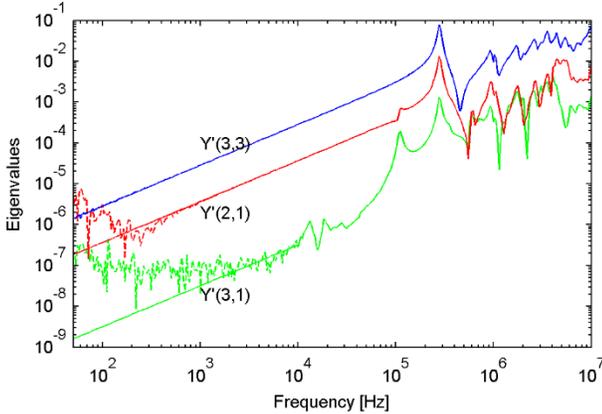

Fig. 8. Common mode measurements. Dashed lines: direct measurement; Solid lines: replacement with capacitive behavior below 10 kHz.

### B. Elimination of Cable Effects

The admittance measurements were performed using coaxial cables of length 3 m. The effect of the measurement cables was removed from **Y** using the transmission line-based cable elimination method described in [11].

Fig. 9 compares the calculated voltage transfer from X1 to Y1 and Y2 (with H1 open), similarly as in Fig. 6. It is observed that the agreement with the directly measured voltage transfers are now greatly improved over the full frequency band. This result is due to 1) the special treatment of the common-mode components in Section IV-A which improves the accuracy at low frequencies, and 2) the cable elimination which improves the accuracy at high frequencies.

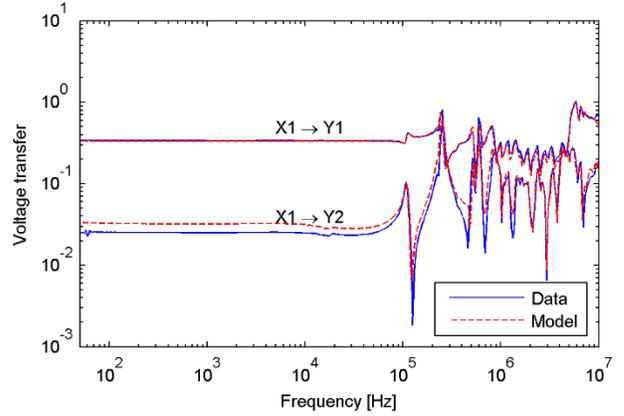

Fig. 9. Voltage transfer from X1 to Y1 and Y2 utilizing common mode measurements and elimination ofd cable effects.

### C. Model Extraction

In order to preserve the small eigenvalues buried in **Y** during the model extraction step, we apply the method of a mode-revealing transformation [12]. Here, an orthogonal similarity transformation **T** is applied (6) such that the small eigenvalues become more visible in the transformed matrix $\tilde{\mathbf{Y}}$ while preserving the passivity properties of **Y**.

$$\mathbf{Y}(\omega) = \mathbf{T}\,\tilde{\mathbf{Y}}(\omega)\,\mathbf{T}^{-1} \qquad (6)$$

The transformed matrix $\tilde{\mathbf{Y}}$ is subjected to pole-residue modeling [13], [14] and passivity enforcement [15]. Fig. 10 shows the elements of $\tilde{\mathbf{Y}}$ and its rational approximation using $N^Y$=80 pole-residue terms. Finally, the inverse transformation of (5) is applied to each residue matrix, giving the rational approximation for **Y** itself (8). Figs. 11 and 12 shows the final result for the rational approximation of **Y** and its eigenvalues, respectively. For use in EMTP-RV, the pole residue model (8) is converted into a state-space model (1).

$$\tilde{\mathbf{Y}}(\omega) \cong \tilde{\mathbf{D}}^Y + \sum_{m=1}^{N^Y} \frac{\tilde{\mathbf{R}}_m^Y}{s - a_m^Y} \qquad (7)$$

$$\mathbf{Y}(\omega) \cong \mathbf{T}(\tilde{\mathbf{D}}^Y + \sum_{m=1}^{N^Y} \frac{\tilde{\mathbf{R}}_m^Y}{s - a_m^Y})\mathbf{T}^{-1} = \mathbf{D}^Y + \sum_{m=1}^{N^Y} \frac{\mathbf{R}_m^Y}{s - a_m^Y} \qquad (8)$$

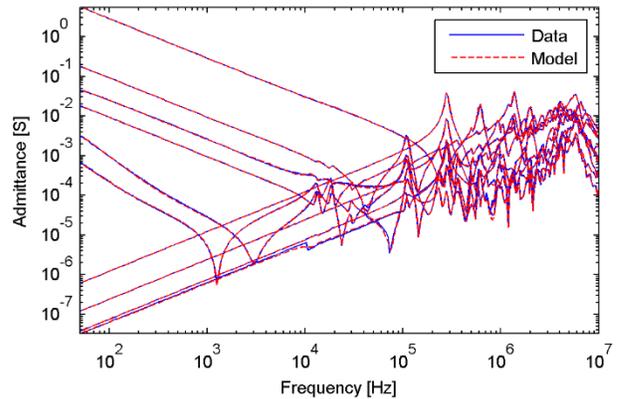

Fig. 10. Fitted elements of $\tilde{\mathbf{Y}}$.

## V. Voltage Transfer Modeling

### A. Measurements

Using two identical voltage probes, the voltage transfer from external terminals 1-4 to internal nodes 5-7 were performed directly on the transformer terminals without the use of coaxial cables. The measurements define the voltage transfer by (2) where the four columns of **H** are subjected to rational fitting by a common pole set of $N^H$=100 poles (9). The pole-residue models are converted [10] into state-space models which are concatenated into a single model (2).

$$\mathbf{h}_i(\omega) \cong \mathbf{r}_{0,i} + \sum_{i=1}^{N^H} \frac{\mathbf{r}_{i,m}}{j\omega - a_{i,m}} \qquad (9)$$

Fig. 11 shows the measured elements along with the rational approximation. It is observed that strong resonances appear in R1 and R5 at around 18 kHz and 1 MHz.

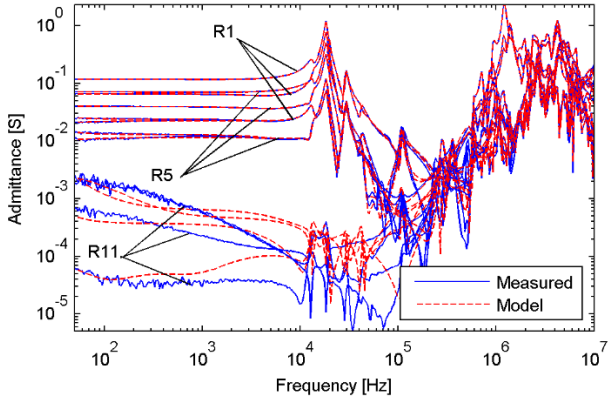

Fig. 11. Rational model of **H** (*N*=100 poles per column).

## VI. Time Domain Validation

### A. Convolution-Based Approach

The final validation of a model should be performed in the time domain since seemingly small model errors in the frequency domain can lead to large errors in the time domain as the simulation time progresses. In this work, we used a slightly different approach based on direct voltage transfer function measurements in the frequency domain with terminal conditions that correspond to the test condition of the accuracy assessment. The transfer functions are measured one-by-one on the terminals using two identical voltage probes. The voltage transfer functions are fitted with a high-order rational function, allowing time domain simulation via recursive convolution [16]. The approach is accurate since there are no matrix manipulations involved. Compared to a direct time domain measurement it has the advantage that the response can be calculated for any input wave shape. In the following subsections these convolution-based simulations are denoted as "Measured" although they have been derived from frequency domain data.

### B. Voltage Application on H1

In this test, the voltage is applied to terminals H1 with H0 and X0 grounded and with X1, Y1 and Y2 open. The voltage transfers from terminal H1 to X1, Y1, Y2, R1, R5 and R11 are measured and fitted with rational functions.

Fig. 12 shows the voltage responses on the external terminals as obtained by the admittance model (1) when applying a 1 Volt lightning impulse wave on H1. The agreement is excellent with only small difference in the oscillation frequency.

Fig. 13 shows the simulated voltage on R1, R5, and R11 as obtained with the voltage transfer model (2) which has the simulated waveforms in Fig. 12 as input. A zoomed view is provided in Fig. 14. Again, a very accurate result has been obtained. Similar accuracy was obtained when applying the voltage excitation on the low-voltage side (X1).

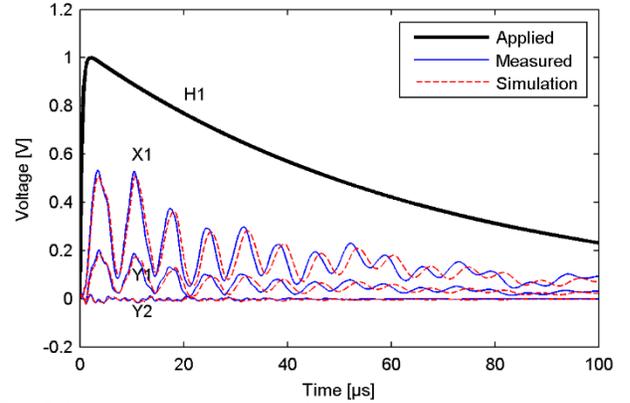

Fig. 12. Voltage response on external nodes.

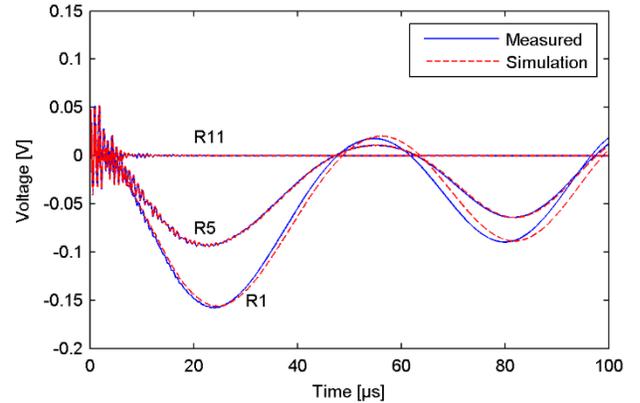

Fig. 13. Voltage response on internal nodes.

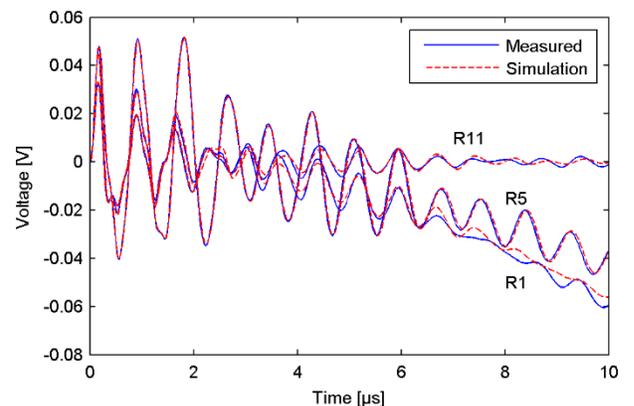

Fig. 14. Voltage response on internal terminals (zoomed view).

## VII. COMPARISON WITH WHITE-BOX MODEL SIMULATION

The transformer manufacturers make use of a detailed models (white-box model) for predicting the internal voltage stresses that arise when applying the standard lightning impulse voltage to external terminals. A lumped-parameter white-box model of the transformer was created based on detailed design information about the transformer and spatial discretization, giving a description in terms of capacitance, inductance and resistance matrices, and internal connections. The resistance values are calculated using the *k*-factor method where $k$ is scaling factor for the resistor elements, see [17] for a description. The modeling conditions are the same as for the actual transformer, i.e. the active part is in tank with no oil and there are no bushings.

Fig. 15 shows the calculated result for the response on X1 with voltage application on H1, which corresponds to one of the traces in Fig. 12. The result with $k=10000$ is seen to give a good agreement for the damping of the dominant natural frequency while usage of $k=0$ (no damping) gives a result which becomes highly inaccurate after a few oscillations. In both cases, the model gives a slightly too high frequency for the oscillation (about 15%).

Fig. 16 shows the calculated result for the response on R1 when applying the voltage on H1, which corresponds to one of the traces in Fig. 13. In this case, usage of $k=10000$ gives a much too high damping of the response while $k=0$ gives a more acceptable result. From the result in Figs. 15 and 16 one can conclude that usage of a single *k*-factor value cannot cover all cases with acceptable accuracy. This result can be explained by the frequency-dependency of the damping as the dominant frequency components in Figs. 15 and 16 are very different (about 100 kHz and 15 kHz, respectively).

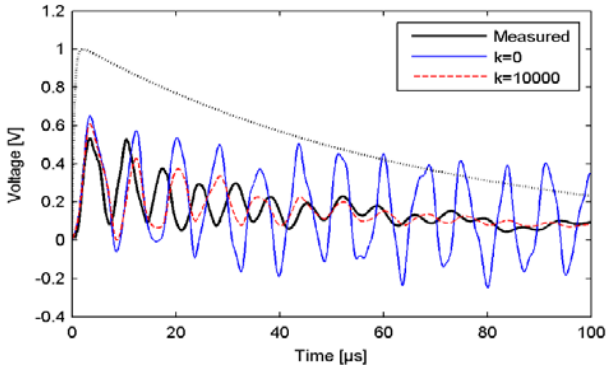

Fig. 15. Voltage application on H1, response on X1 by white-box model.

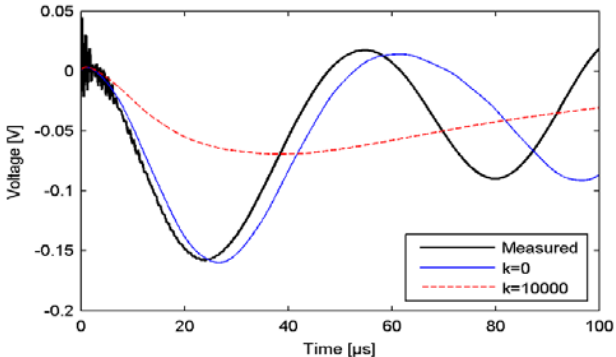

Fig. 16. Voltage application on H1, response on R1 by white-box model.

## VIII. IMPACT OF TAP SETTING ON TERMINAL BEHAVIOR

The selected tap setting will impact the overvoltages that occur in the regulating winding. In addition, the tap setting also affects the transferred overvoltages between the windings. The transferred voltage was investigated by measuring the admittance matrix with respect to terminals H1 and X1 with the tertiary terminals (Y1, Y2) open. The elements of the associated 2×2 admittance matrix are shown in Fig. 17, with nodes 1 and 2 denoting H1 and X1, respectively. The elements are shown for the transformer tap setting in position Max, Nom+, Nom- and Min. It can be seen that a noticeable difference exists between the tap positions at frequencies below 30 kHz. The difference between the two mid-positions Nom+ and Nom- is however very small. Above 30 kHz, the differences between all curves are small until 1 MHz where noticeable differences can again be observed.

From the admittance matrices, the voltage transfer function X1 to H1 was calculated, see Fig. 18. It can be observed that the tap setting greatly affects the first resonance peak at about 7 kHz. The peak is substantially reduced and shifted towards higher frequencies as the tap position is moved from Max to Mid (Nom+/Nom-) and Min. These results imply that a general black-box model should be created for more than a single tap position to properly represent the low frequency resonance. The change to the low frequency resonance is quantified in Table III.

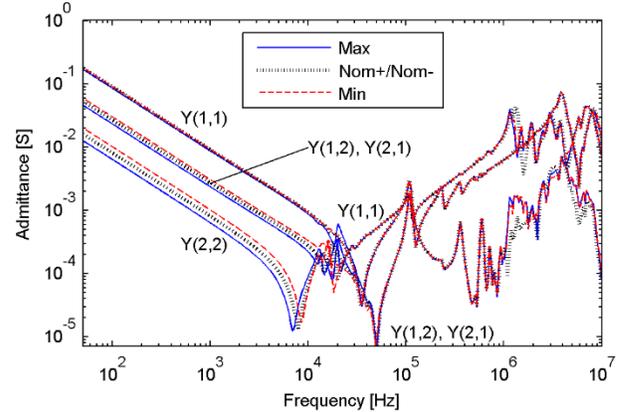

Fig. 17. Measured terminal admittance with respect to H1 and X1, with tertiary terminals (Y1, Y2) open. Impact of tap setting.

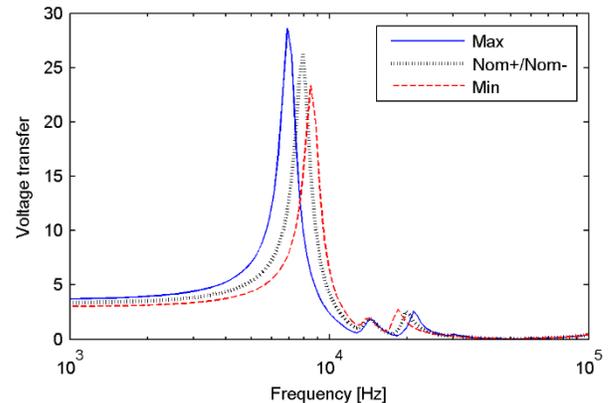

Fig. 18. Calculated voltage transfer from X1 to H1.

TABLE III LOW-FREQUENCY RESONANCE IN VOLTAGE TRANSFER FROM X1 TO H1

| Position | $f_0$ [kHz] | $|H_0|$ |
|---|---|---|
| Max | 6.9 | 29 |
| Mid | 7.9 | 26 |
| Min | 8.5 | 23 |

## IX. APPLICATION EXAMPLE

We return to the complete transformer model in tap Nom+ and demonstrate its potential application in a transient simulation study. Three transformer units are connected into a three-phase bank as shown in Fig. 19 by connecting the tertiary windings into delta. The transformer is fed from a substation via a cable as shown in Fig. 20. The cable is modeled as a three lossless single-phase lines with propagation characteristics as shown in Table V, and with mutual coupling between phases neglected. The system overhead lines (OHL) are single-circuit three-phase lines over an earth with resistivity 100 Ω·m. The conductor data are given in Table IV by position ($x$, $y$) diameter ($d$) and DC resistance ($R_{dc}$). The overhead lines are modeled in EMTP-RV as multi-conductor lines with inclusion of frequency-dependent effects.

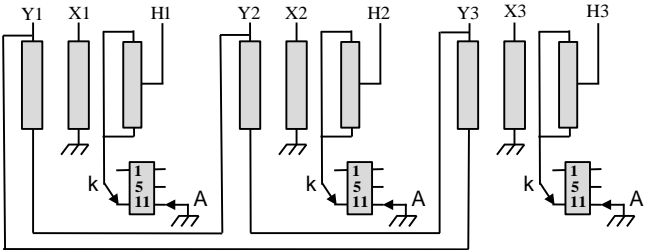

Fig. 19. Connecting three 1-phase transformers into a 3-phase bank.

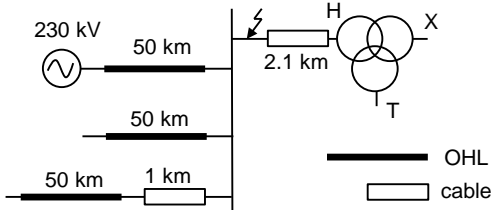

Fig. 20. Ground fault initiation in phase 1 of transformer feeding cable.

TABLE IV OVERHEAD LINE CONDUCTOR PARAMETERS

| Conductor no. | 1 | 2 | 3 | 4 | 5 |
|---|---|---|---|---|---|
| $x$ [m] | −7.6 | 0 | 7.6 | −5 | 5 |
| $y$ [m] | 19.5 | 19.5 | 19.5 | 24.4 | 24.4 |
| $d$ [cm] | 3.51 | 3.51 | 3.51 | 1.94 | 1.94 |
| $R_{dc}$ [Ω/km] | 0.045 | 0.045 | 0.045 | 0.151 | 0.151 |

TABLE V CABLE PROPAGATION PARAMETERS

| Characteristic impedance, $Z_C$ | Velocity, $v$ |
|---|---|
| 30 Ω | 160 m/μs |

An ideal ground fault occurs in phase 1 in the substation, at voltage maximum, which is represented in the simulation by a closing switch. The ground fault initiation results in a voltage oscillation in phase 1 on the transformer HV side. The dominant frequency component results from quarter-wave oscillation in the line stub. The frequency is estimated by (10) to be 19 kHz with $l$=2.1 km and $v$=160 m/μs. This oscillation frequency coincides with the resonance peak in voltage transfer from the HV terminal to the internal points in Fig. 11.

$$f_{\lambda/4} = \frac{v}{4l} \quad (10)$$

Fig. 21 shows the simulated voltage waveforms on the transformer external terminals and at the internal points (R1, R5, R11), in the faulted phase. It is observed that the internal voltages undergo a resonant build-up, giving a peak value of about 200 kV. For comparison, the ditto peak voltage during the lightning impulse test (all terminals grounded except H1) with $U_{BIL}$=1050 kV peak voltage was found to reach only 182 kV.

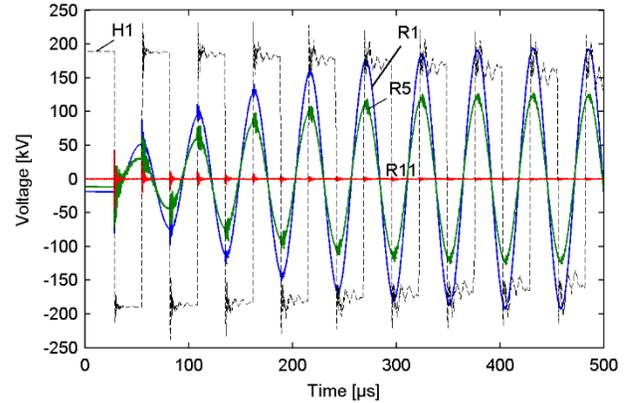

Fig. 21. Voltage on external terminals and internal points in phase 1.

The damping of the impinging voltage on the HV terminal is greatly increased when replacing the cable with an overhead line due to its higher characteristic impedance. To see this, the 2.1 km cable between the bus and the transformer is replaced with a 4 km overhead line. The quarter-wave resonance frequency of the OHL is the same as for the 2.1 km overhead line, i.e. 19 kHz. The overvoltage in R1 in relation to the peak value of the 60 Hz stationary voltage is listed in Table VI when the ground fault occurs on either the cable end (Fig. 21) or one the OHL end.

TABLE VI OVERVOLTAGE IN R1 FROM GROUND FAULT INITIATION

| Feeding line | Overvoltage factor in R1 |
|---|---|
| Underground cable (2.1 km) | 10.2 |
| Overhead line (4 km) | 4.4 |

## X. DISCUSSION

The measurements were performed with the transformer in tank, but without oil. Measurements in oil are possible but such approach is less attractive for reasons related to time consumption and cost. The impact of the missing oil is that the partial capacitances between transformer internal parts is too low, which is affects both admittance and voltage transfer functions. In general, reducing the capacitances will shift resonances towards high frequencies. This impact of the missing oil should be clarified by further studies. The measurements were further performed without presence of the bushings. The bushings can however be represented by lumped shunt capacitances that can be added to the model.

## XI. CONCLUSION

This work has demonstrated that overvoltages in the regulating winding can be simulated using a black-box model approach which combines a terminal equivalent with a voltage transfer model. The approach offers excellent accuracy and compatibility with EMTP-type simulation programs.

The presence of an ungrounded tertiary winding causes error magnifications at lower frequencies. This problem is overcome by explicitly representing the common mode components during measurement and model extraction.

Usage of the model in a transient simulation study shows that internal voltages in the regulating winding can in resonant conditions exceed the voltage that occurs in the lightning impulse test. The model can therefore be a valuable tool for identifying network conditions where such resonant voltage buildup are likely to occur.

The tap setting can substantially impact the terminal admittance matrix below 10 kHz, thereby impacting the lower-frequency resonance frequencies. Black-box terminal models should therefore be provided for several tap settings, e.g. Min. Mid and Max positions.

## XII. ACKNOWLEDGMENT

The authors thank their colleagues in CIGRE JWG A2/C4.52 for their support and many useful discussions.

## XIV. BIOGRAPHIES

**Bjørn Gustavsen** (M'94–SM'2003–F'2014) was born in Norway in 1965. He received the M.Sc. degree and the Dr.Ing. degree in Electrical Engineering from the Norwegian Institute of Technology (NTH) in Trondheim, Norway, in 1989 and 1993, respectively. Since 1994 he has been working at SINTEF Energy Research where he is currently a Chief Research Scientist. His interests include simulation of electromagnetic transients and modeling of frequency dependent effects. He spent 1996 as a Visiting Researcher at the University of Toronto, Canada, and the summer of 1998 at the Manitoba HVDC Research Centre, Winnipeg, Canada. He was a Marie Curie Fellow at the University of Stuttgart, Germany, August 2001–August 2002. He is convener of CIGRE JWG A2/C4.52 "High-frequency transformer and reactor models for network studies".

**Álvaro Portillo** (M'84–SM'2001) was born in Uruguay in 1954. He graduated in Electrical Engineering in the Uruguayan Republic University in 1979. He worked in the Uruguayan electrical utility (UTE) up to 1985 in activities related with acceptance, installation and maintenance of power transformers. From 1985 to 1999 he worked in MAK S.A. (Uruguayan manufacturer of transformers), from 2000 to 2007 as consultant in TRAFO (Brazilian manufacturer of transformers) and from 2007 up today as consultant, in developing software tools for transformers design, at WEG (Brazilian manufacturer of transformers). He is a professor at the Uruguayan Republic University since 1977, now responsible for post-graduation courses about transformers (specification, design, design review, operation, maintenance, repairs, etc.). He is task force leader within CIGRE JWG A2/C4.52 "High-frequency transformer and reactor models for network studies".

**Rodrigo Ronchi** was born in Brazil in 1980. He graduated with P.E. degree in Electrical Engineering in Fundacao Universidade Regional de Blumenau - FURB in 2008 and 2010, respectively. He worked at ABB in Blumenau up to 2005 in Engineering Department with distribution transformer design. Since 2005 he has been working at WEG Transformadores (Brazil and México) with power transformer design. He is member of CIGRE JWG A2/C4.52 "High-frequency transformer and reactor models for network studies".

**Asgeir Mjelve** was born in Oslo, Norway, in 1959. He received the B.Sc. degree in electrical engineering from Østfold University College, Norway, in 1980. Since 1982, he has been with Hafslund Nett, the electricity utility for the city of Oslo and surrounding areas and the largest distribution grid company in Norway. He has been working primarily with issues related to planning, design, and maintenance of substations and is also responsible for the R&D activities in Hafslund Nett. He is member of CIGRE Transformers Study Committee.